\documentclass[12pt,times,twocolumn]{aastex631}

\journalinfo{Accepted for publication in ApJL}

\usepackage{xspace}

\newcommand{\Change}[1]{{\bf #1}}
\renewcommand{\Change}[1]{#1}
\newcommand{\CChange}[1]{{\bf #1}}
\renewcommand{\CChange}[1]{#1}

\newcommand{\mang}   {\,\mathrm{{\mbox{\AA}}}\xspace}
\newcommand{\kev}    {\,\mathrm{keV}}
\newcommand{\ks}     {\,\mathrm{ks}}
\newcommand{\pc}     {\,\mathrm{pc}}
\newcommand{\mk}     {\,\mathrm{MK}}
\newcommand{\msun}   {\,M_\odot}
\newcommand{\xray}   {{X-ray}\xspace}
\newcommand{\chan}   {{\it Chandra}\xspace}
\newcommand{\hetg}   {{HETG}\xspace}
\newcommand{\xmm}    {{\it XMM-Newton}\xspace}
\newcommand{\swift}  {{\it Swift}\xspace}
\newcommand{\nustar} {{\it NuSTAR}\xspace}

\newcounter{ion} \newcommand{\eli}[2]{\setcounter{ion}{#2}#1{~\sc\roman{ion}}}

\newcommand{\mone}    {^{-1}}
\newcommand{\mtwo}    {^{-2}}
\newcommand{\mthree}  {^{-3}}
\newcommand{\cmmtwo}  {\,\mathrm{cm\mtwo}}
\newcommand{\cmmthree}{\,\mathrm{cm\mthree}}
\newcommand{\eflux}   {\,\mathrm{erg\,cm\mtwo\,s\mone}}
\newcommand{\kms}     {\,\mathrm{km\,s\mone}}
\newcommand{\lum}     {\,\mathrm{erg\,s\mone}}
\newcommand{\mdot}    {\msun\,\mathrm{yr}\mone}

\newcommand{\pfluxs}  {\,\mathrm{phot\,cm\mtwo\,s\mone}}

\newcommand{\piaqr}{{$\pi\,$Aqr}\xspace}
\newcommand{\gcas}{{$\gamma\,$Cas}\xspace}

\newcommand{\mki}{
  Kavli Institute for Astrophysics and Space Research, 
  Massachusetts Institute of Technology , 77 Massachusetts  Ave., 
  Cambridge, MA 02139, USA
}

\shorttitle{High Resolution X-ray Observations of $\pi\,$Aqr}
\shortauthors{Huenemoerder et al.}

\graphicspath{{./}{figures/}}

\begin{document}

\title{Chandra HETG X-ray Spectra and Variability of \piaqr, a \gcas-type Be star.}

\author[0000-0002-3860-6230]{David P.\ Huenemoerder}\thanks{dph@mit.edu}\affiliation{\mki}

\author[0000-0002-1131-3059]{Pragati Pradhan}
 \affiliation{Embry Riddle Aeronautical University, Department of
 Physics \& Astronomy, 3700 Willow Creek Road Prescott, AZ 86301, USA}

\author[0000-0002-5769-8441]{Claude R. Canizares} \affiliation{\mki}

\author[0000-0003-2602-6703]{Sean Gunderson} \affiliation{\mki}

\author[0000-0002-7204-5502]{Richard Ignace} \affiliation{Department
 of Physics \& Astronomy, East Tennessee State University, Johnson City, TN 37614 USA}

\author[0000-0003-3298-7455]{Joy~S.\ Nichols}
\affiliation{Harvard \& Smithsonian Center for Astrophysics, 60 Garden St., Cambridge, MA 02138, USA}

\author[0000-0002-6737-538X]{A.~M.~T. Pollock}
\affiliation{Department of Physics and Astronomy, University of Sheffield, Hounsfield Road, Sheffield S3 7RH, UK}

\author{Norbert S.\ Schulz}\affiliation{\mki}

\author[0000-0003-1898-4223]{Dustin K. Swarm}
\affiliation{Department of Physics and Astronomy, University of Iowa, Iowa City, IA, USA}

\author[0000-0002-5967-5163]{Jos\'e M. Torrej\'{o}n}
\affiliation{Instituto Universitario de F\'{\i}sica Aplicada a las Ciencias y las Tecnlog\'{\i}as, Universidad de Alicante, E-
03690 Alicante, Spain}


\begin{abstract}
High-resolution X-ray spectra of \piaqr, a $\gamma\,$Cas-type star,
obtained with the \chan/\hetg grating spectrometer, revealed emission
lines of H-like ions of Mg, Si, S, and Fe, a strong, hard continuum,
and a lack of He-like ions, indicating the presence of very hot
thermal plasma.  The X-ray light curve showed significant
fluctuations, with coherent variability at period of about $3400\,$s
in one observation.  The hardness ratio was relatively constant except
for one observation in which the spectrum was much harder and more
absorbed.  We interpret the X-ray emission as arising from accretion
onto the secondary, which is likely a magnetic white dwarf, an
intermediate polar system.
\end{abstract}




\section{Introduction}\label{sec:intro}

\piaqr was recently identified as a \gcas-type star by
\citet{naze:al:2017}.  The \gcas-types are O/Be stars which in X-rays
have unusually hard spectra, with characteristic temperatures
exceeding $100\mk$ ($8.6\kev$), yet appear to be thermal due to
presence of \eli{Fe}{25} and \eli{Fe}{26} emission lines. Given their
thermal spectrum and low X-ray luminosities, their nature as accreting
X-ray binaries containing a neutron star or a black hole companion is
ruled out. Several \gcas types are known to be some form of binary
including \piaqr \citep{bjorkman:al:2002}: it is the nature of the
companion which makes them of special interest.
\citet{postnov:al:2017} proposed a neutron star where direct accretion
is impeded magneto-centrifugally, the so called propeller
state. \citet{langer:al:2020} gave an account of possible evolutionary
states, with preference for stripped-core companions, either SdO or
He-stars. \citet{gies:al:2023} argued that SdO companions would be
bright enough to see optically, so they prefer a white dwarf companion
hypothesis.  \citet{tsujimoto:al:2018}, based on X-ray spectral
modeling of two \gcas-type stars, also prefer a white dwarf companion.
In \xmm spectra of \piaqr, \citet{naze:al:2017} found a plasma
temperature of about $10\kev$ ($116\mk$) from isothermal fits of
thermal spectra, and found no significant emission line features in
the lower-energy RGS spectral band.  In \swift data,
\citet{naze:rauw:smith:2019} had similar results requiring an
isothermal plasma with $kT =15\kev$ ($T=174\mk$), and local absorption
of about $N_\mathrm{H} = 0.5\times10^{22}\cmmtwo$.
\citet{tsujimoto:al:2023} modeled \nustar and \xmm spectra of \piaqr
and found a maximum temperature of $19\kev$ ($220\mk$) from a cooling
flow model \citep{pandel:al:2005}, concluding that \piaqr could host
either a magnetic or non-magnetic white dwarf.
\Change{The interferometric survey of Be stars by
  \citet{klement:rivinius:al:2024} included several \gcas-type stars
  with \xray emission.  They failed to detect any companions, and
  excluded SdO and main-sequence companions for \piaqr, leaving
  a white dwarf as the most viable secondary and
  source of \xray emission.}

Here we present X-ray spectral and temporal properties of \piaqr
revealed by recent \chan/\hetg observations.  \S~\ref{sec:obs} details
the observations and reduction of the {\em Chandra} HETG data.  The
spectra are described and interpreted through spectral modeling in
\S~\ref{sec:spec}.  Finally in \S~\ref{sec:var} we comment on
variability observed in our dataset, in particular a notable hardening
of the spectra arising from a jump in absorbing column density during
the last of our observational sets.

\section{Observations and Processing} \label{sec:obs}

We obtained \chan/\hetg data for \piaqr in six
observations made during August to October, 2022, for a total exposure
of $101\ks$.  Details are provided in Table~\ref{tbl:obsinfo}.\footnote{Data are available at
\url{https://doi.org/10.25574/ObsID}, where $ObsID$ is to be replaced
with the values from Table~\ref{tbl:obsinfo}.}
\citet{canizares:al:2005} gave a description of the instrument. Data
were processed with CIAO \citep{CIAO:2006} following standard
procedures, except for use of a narrower spatial region defining the
HEG and MEG loci. This removes ambiguity between the two grating types
at the shortest wavelengths, and can improve flux determination below
$\sim1.8\mang$ in particular for hard sources where standard regions
overlap.  Spectra were analyzed using the ISIS package
\citep{houck2000} in conjunction with AtomDB \citep{foster2012}.  For
timing analysis, we used the SITAR ISIS package\footnote{For a
description of SITAR routines, see
\url{https://space.mit.edu/cxc/analysis/SITAR/}.}, as distributed with
ISISscripts.\footnote{The ISISscripts package is available from
\url{http://www.sternwarte.uni-erlangen.de/isis/}.}

%
\begin{deluxetable}{cccc}
\tablecaption{\piaqr Observational Information\label{tbl:obsinfo}}
\tablewidth{0pt}
\tablehead{
\colhead{ObsID}& 
\colhead{DATE-OBS}&
\colhead{Exposure} & 
\colhead{$\phi_\mathrm{orb}$} \\
 &
 \colhead{[\chan start date]}&
 \colhead{[$\ks$]}&
}
\startdata
   \dataset[26079]{\doi{10.25574/26079}}&  2022-08-23T15:37:02&          9.93&    0.40 \\
   \dataset[27269]{\doi{10.25574/27269}}&  2022-08-27T04:24:31&          9.99&    0.44 \\
   \dataset[26080]{\doi{10.25574/26080}}&  2022-09-05T00:58:47&         29.67&    0.54 \\
   \dataset[26001]{\doi{10.25574/26001}}&  2022-09-13T00:14:35&         19.80&    0.64 \\
   \dataset[27412]{\doi{10.25574/27412}}&  2022-09-14T06:16:23&         21.78&    0.65 \\
   \dataset[27325]{\doi{10.25574/27325}}&  2022-10-30T09:44:35&          9.76&    0.20 \\
\enddata
\tablecomments{\raggedright The orbital phase is defined as $0.0$ for
  the maximum radial velocity (redshift) of the secondary, using the
  error-weighted-mean ephermeris of \citet{bjorkman:al:2002}: $T_0 =
  2450274.84\,\mathrm{JD}$, $P=84.132\,\mathrm{day}$.}
\end{deluxetable}

\section{Spectrum and Modeling}\label{sec:spec}

\begin{figure*}[ht]
  \centering\leavevmode
  \includegraphics*[width=0.68\textwidth]{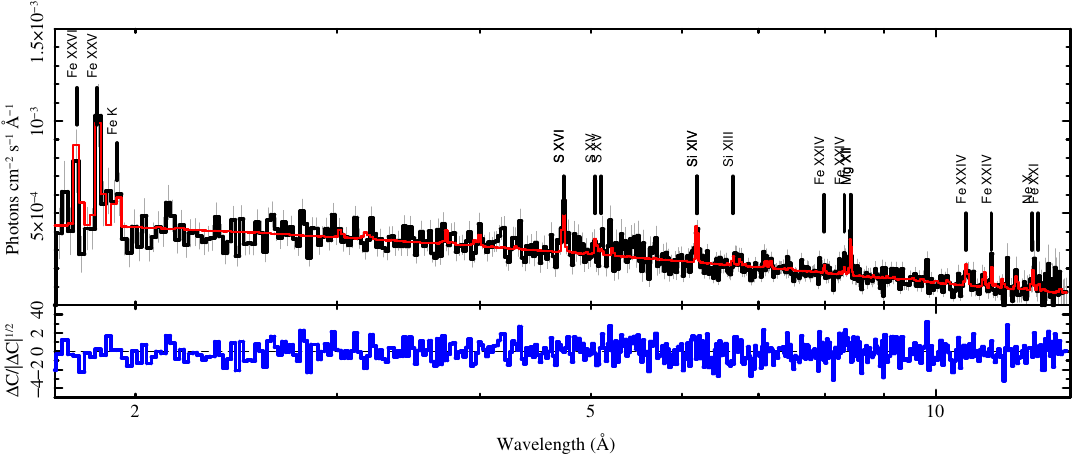}
  \includegraphics*[width=0.29\textwidth]{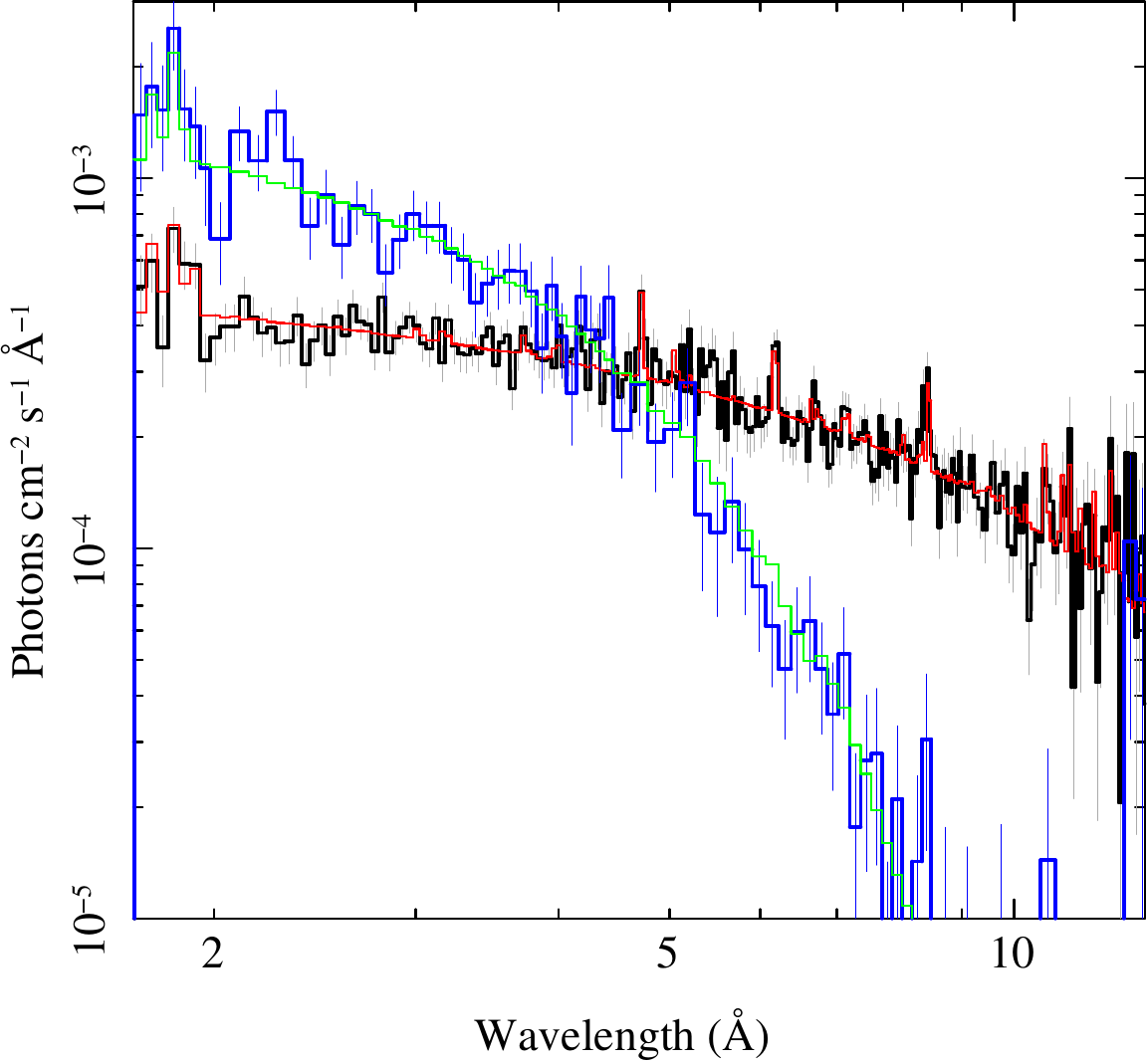}
 \caption{Left: HETG
  spectrum of \piaqr\ in black, which is the merged HEG and MEG first
  orders, excluding the high hardness ratio observation (27325).  In
  red is a model using three APEC components plus a Gaussian for the
  putative Fe~K fluorescence.  On the right, we compare the high
  hardness ratio state spectrum (blue) to the rest (black and
  red).} \label{fig:spec}
\end{figure*}

We analyzed the spectrum in two parts because one observation, with
about $10\ks$ of the $100\ks$ total exposure, had a very different
character in terms of hardness (see \S~\ref{sec:var} for variability details).  We show the
low hardness ratio state \hetg\ spectrum in the left side of
Figure~\ref{fig:spec}, where strong \eli{Fe}{25}
($1.85\mang$) and \eli{Fe}{26} ($1.78\mang$) emission lines are seen, with a
shoulder on the red side of the He-like line that may be Fe~K$\alpha$
fluorescence.  The rest of the spectrum clearly shows the
H-Ly$\alpha$-like emission lines from \eli{S}{16} ($4.73\mang$),
\eli{Si}{14} ($6.18\mang$), and \eli{Mg}{12} ($8.42\mang$) on top of a
strong continuum.  There may be some \eli{Fe}{24} emission lines near
\eli{Mg}{12} and between $10$--$13\mang$. The He-Ly$\alpha$-like
emission lines from S, Si, and Mg are conspicuous in their absence.
These characteristics point to very high-temperature plasmas; the
H-like lines, except for Fe, are weak because the plasma is dominated
by temperatures well above the peaks of the emissivity functions, on
the long tail of the H-like emissivities and well beyond the
significant region of He-like emissivity.  In an isothermal plasma, to
obtain a limit in the \eli{Si}{13} to \eli{Si}{14} flux less than the
observed ratio of $<0.10$ requires a temperature greater than $42\mk$,
and to obtain the observed ratio of \eli{Fe}{26} to \eli{Fe}{25}
requires a temperature of at least $70\mk$.

To characterize the spectrum more systematically, we have fit a
few-temperature model using AtomDB emissivities.  For such a hot
plasma, the continuum shape provides limited leverage on the
temperature, since above about $100\mk$, the turnover of the thermal
bremsstrahlung occurs below $2\mang$, and the shape above that is
largely flat and featureless.  We find that three temperatures are
sufficient to describe the spectrum, to which we added a Gaussian for
the possible Fe K$\alpha$ fluorescence.  Each temperature component
shared a common Doppler shift and ``turbulent'' broadening term for
any Gaussian intrinsic width. Instrumental broadening is handled by
forward-folding through the calibrated responses, and common
elemental abundances, for which Mg, Si, S, and Fe were free to vary.
However, for few-temperature plasmas, abundances are degenerate with
temperature since there are two ways to change a model line flux: vary
the temperature, or vary the abundance. Should the temperature not
be accurate, the abundance will be not be accurate as well. We also include a single
absorption term, however note that the foreground interstellar absorption is small in this
band ($\sim0.03\times10^{22}\cmmtwo$, \citet{gudennavar:al:2012}).

The fit converges on a very hot component to produce enough
\eli{Fe}{26} and to provide the correct shape between $1.7\mang$ an
$3\mang$.  However, the \eli{Fe}{26} flux is rather uncertain by about
40\%, so anything over $\sim 200\mk$ would be adequate.  A component
of about $40\mk$ is required to give the observed \eli{Fe}{25} flux,
some \eli{Fe}{26}, and most of the \eli{S}{16} emission, plus
\eli{Si}{14}, \eli{Mg}{12}, and \eli{Fe}{24}.  The third component, at
$10\mk$ also contributes to the H-like series, and to the 
\eli{Fe}{24} ionization, but this component has the lowest normalization. However, if
larger, it would contribute to \eli{Si}{13}, which is weak or absent.
The emission measure, for a distance of $286\pc$
\citep{gaia:2016,gaia:2018} is $1.0\times10^{55}\cmmthree$.  The
absorption required is $N_\mathrm{H} \sim 0.4\times 10^{22}$, which has a transmission
factor of about $0.3$ at $13\mang$.  The absorbed model flux in the
$0.5$--$10\kev$ band is $1.5\times10^{-11}\eflux$, and the X-ray
luminosity is $1.8\times10^{32}\lum$.  The
lines required some broadening beyond instrumental, with a turbulent
velocity term of $\sim 400\kms$, which corresponds to \eli{S}{16},
\eli{Si}{14} and \eli{Mg}{12} full-widths, half-maxima ($FWHM$s) of
about $900\kms$ (for $T=40\mk$).  The Doppler shift from the global fit was $+64\pm66\kms$, 
without correction for the line-of-sight velocities at these epochs, which is less than $7\kms$.  
The range is incommensurate with the \citet{bjorkman:al:2002} solution for the secondary at these 
phases of about $-70\kms$.   Plasma model details are given in Table~\ref{tbl:plasmafit}.

We have also fit the few emission lines with intrinsic Gaussian
profiles, folded through the instrument response, in order to
determine fluxes, velocity offsets, and widths.  The features with
signal-to-noise ratio $>3$ and which have good resolution are
\eli{Si}{14} and \eli{Mg}{12}.  These have $FWHM\sim800\kms$,
consistent with the plasma model determination.  Their Doppler shifts
are consistent with $0\kms$.  Line fit details are given in
Table~\ref{tbl:linefit}.

The right panel of
Figure~\ref{fig:spec} shows the high hardness state spectrum compared to the low hardness state. We see that the high hardness 
state is brighter below $4\mang$, but
much more absorbed.  We fit this hard state spectrum with a two temperature model
since less information is available from emission lines - only
\eli{Fe}{25}, being so heavily absorbed, and having only $10\ks$
exposure. The plasma is much hotter, has higher flux, and much greater
absorption.  Details are given in Table~\ref{tbl:plasmafit}

\begin{deluxetable}{c|cc|cc|c}
\tablecaption{\piaqr Plasma Model Fit Parameters\label{tbl:plasmafit}}
\tablewidth{0pt}
\tablehead{
\colhead{Parameter} & 
\colhead{Value} & 
\colhead{$\sigma$}&
\colhead{Value} & 
\colhead{$\sigma$}&
\colhead{Unit}\\
& \multicolumn{2}{c|}{Normal HR}
& \multicolumn{2}{c|}{High HR}&
}
\startdata
$T_1$& $270$&  $94$& $>520$&   & $\mk$\\
$T_2$&  $44$&  $11$& $70$& $20$& "\\
$T_3$&  $13$&  $3$&      &     & "\\
$norm_1$& $7.3$& $0.8$& $21.1$& $3.6$& $10^{-11}\mathrm{cm^{-5}}$\\
$norm_2$& $2.7$& $1.0$& $6.0$& $2.8$&  "\\
$norm_3$& $0.4$& $0.3$& &&             "\\
$v_\mathrm{turb}$ & $404$& $169$& && $\kms$\\
$\Delta v$ &$64$& $66$& && "\\
$A$(Mg) &$1.2$& $0.3$& && Relative to Solar\\
$A$(Si) &$0.7$& $0.2$& && "\\
$A$(S)  &$1.3$& $0.4$& && "\\
$A$(Fe) &$0.6$& $0.1$& && "\\
$N_\mathrm{H}$& $0.40$& $0.05$& $4.7$& $0.2$& $10^{22}\cmmtwo$\\
$EM$ &$1.0$& $0.1$& $2.6$& $0.4$&         $10^{55}\cmmthree$ \\
$f_\mathrm{x}$& $1.50$&   & $2.54$& &      $10^{-11}\eflux$\\
$L_\mathrm{x}$& $1.81$ && $4.38$& &        $10^{32}\lum$ \\
\enddata
\tablecomments{``Normal HR'' refers to all observations except 27325,
  and ``High HR'' is for 27325 in the high hardness-ratio state.  Flux
  is integrated from the model over the $0.5$--$10\kev$ band.  The
  emission measure used a distance of $D=286\pc$, and is related the
  the normalizations by a factor of $10^{11} 4\pi D^2$.  Relative
  abundances are referenced to \citet{anders:89}. Uncertainties
  are $1\sigma$.}
\end{deluxetable}

\begin{deluxetable}{llrrrrrr}
\tablecaption{\piaqr Line Fit Parameters\label{tbl:linefit}}
\tablewidth{0pt}
\tablehead{
\colhead{Line}& 
\colhead{$\lambda_0$}&
\colhead{$\Delta v$} & 
\colhead{$\sigma$}&
\colhead{$10^6\times$Flux}& 
\colhead{$\sigma$}&
\colhead{{\footnotesize$FWHM$}} & 
\colhead{$\sigma$}\\
 &
 \colhead{[$\mang$]}&
 \multicolumn{2}{c}{[$\kms$]}&
 \multicolumn{2}{c}{[$\pfluxs$]}&
 \multicolumn{2}{c}{[$\kms$]}
}
\startdata
\eli{Fe}{26}&
$1.7798$&
\nodata&
\nodata&
$8.5$&
$3.4$&
$0.0$&
$1320$\\
\eli{Fe}{25}&
$1.8554$&
$-48$&
$276$&
$20.7$&
$4.8$&
$2372$&
$1101$\\
FeK$\alpha$&
$1.94$&
$-950$&
$773$&
$4.5$&
$2.4$&
\nodata&
\nodata\\
\eli{S}{16}&
$4.7301$&
$-240$&
$630$&
$5.3$&
$2.1$&
\nodata&
\nodata\\
\eli{Si}{14}&
$6.1831$&
$92$&
$123$&
$6.0$&
$1.5$&
$778$&
$582$\\
\eli{Si}{13}(r)&
$6.6479$&
\nodata&
\nodata&
$<1.2$&
\nodata&
\nodata&
\nodata\\
\eli{Mg}{12}&
$8.4219$&
$28$&
$71$&
$9.9$&
$1.7$&
$840$&
$280$\\
\enddata
\tablecomments{\raggedright Wavelengths for \eli{Fe}{26} and
  \eli{Fe}{25} are weighted means over the components in each group;
  each was fit with a single Gaussian model.  The \eli{Fe}{26} offset
  from \eli{Fe}{25} was frozen. The width of FeK$\alpha$ was tied to
  that of \eli{Fe}{26}. The \eli{S}{16} width was tied to that of
  \eli{Si}{14}, since the former has a fairly low signal-to-noise
  ratio.  The flux limit on the \eli{Si}{13} resonance line was
  computed with a frozen position and a width frozen at the value for
  \eli{Si}{14}. Uncertainties are for $68\%$ confidence intervals.}
\end{deluxetable}

\section{Variability}\label{sec:var}

It is well known that \gcas-type stars are variable in optical and
X-ray bands
\citep{naze:pigulski:al:2020,naze:rauw:al:2020,naze:motch:al:2020}.
Optical variability can be highly periodic and is generally attributed
to the B-star's rotation or pulsations.  X-ray fluctuations can exceed
factors of two or three in short times, behavior often referred
to as ``flaring'' (but not like coronal flares
in the Sun and late-type stars, which have increases in both
flux and temperature and are strongly tied to
stellar magnetism).

\begin{figure*}
  \centering\leavevmode
  \includegraphics*[width=0.64\textwidth]{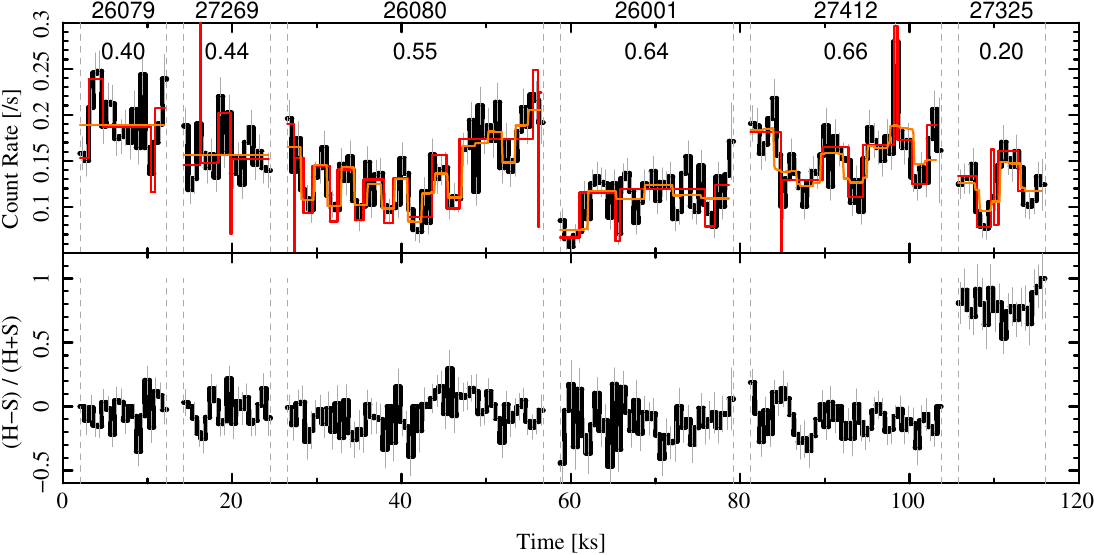}  \hfill
  \includegraphics*[width=0.33\textwidth,viewport=0 -10 550 530]{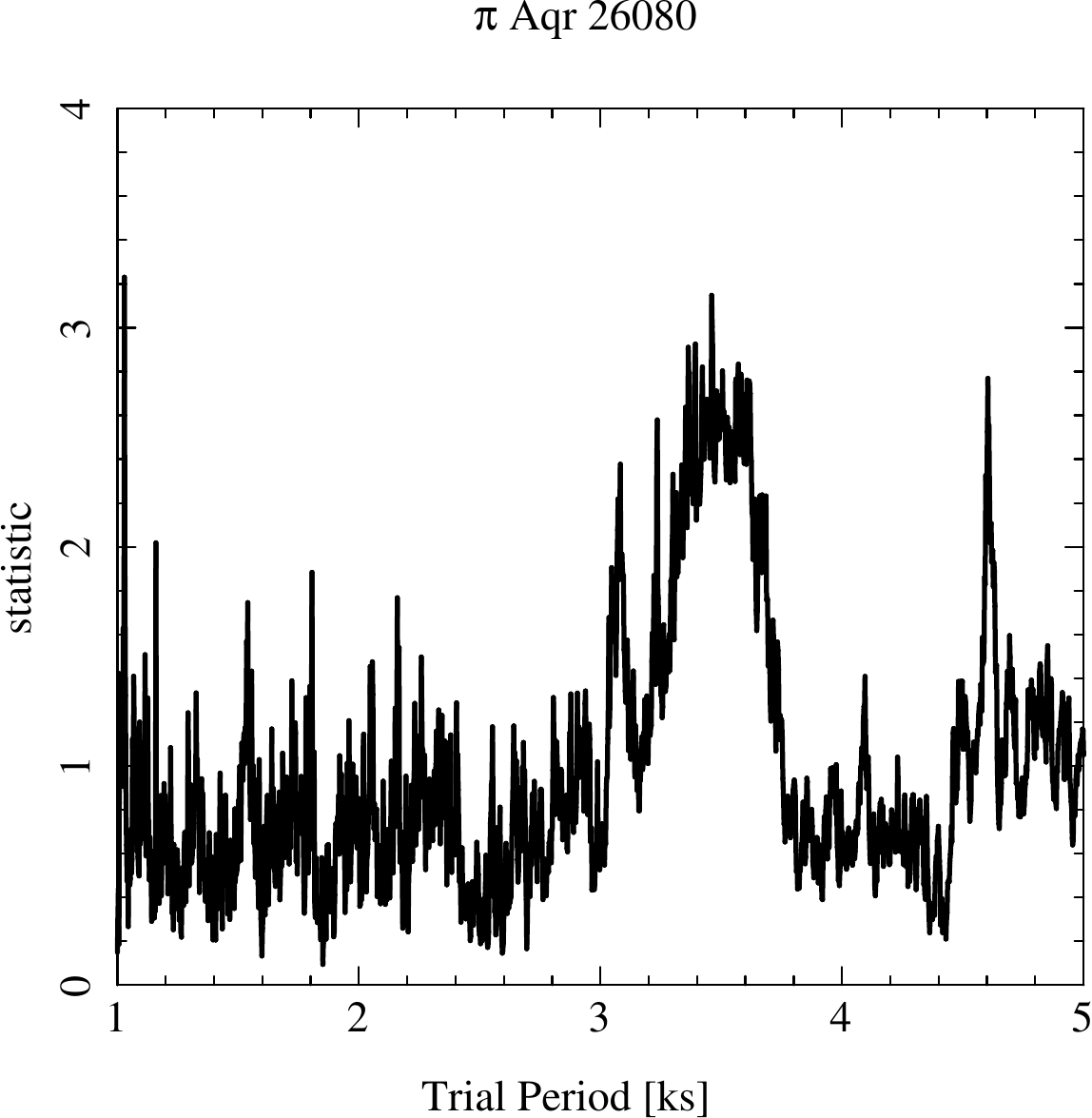} 
  \caption{Left top: The X-ray light curve of \piaqr for all
  observations for the $1.7$--$13\mang$ band, concatenated in time
  order.  The vertical dashed lines mark arbitrary length gaps between
  observations.  The black curve shows the count rate in dispersed
  first-order events.  The red and orange curves show the results of
  Bayesian blocks \citep{scargle:al:2006, scargle:al:2013} and
  Gregory-Loredo \citep{gregory:loredo:1992} algorithms, confirming
  the statistically significant variability.  Left bottom: the
  count-rate hardness ratio, $(H-S)/(H+S)$, where $H$ is the ``hard''
  rate in the $1.7$--$5.2\mang$ band, and $S$ in $5.2$--$13\mang$ for
  ``soft''.  The last observation (27325) does not stand out in the
  total rate curve, but does in hardness.  Labels at the top give the
  observation IDs with orbital phases below them.  Right: epoch
  folding algorithm \citep{davies:1990} ``$L$''-statistic for obsid
  26080 showing coherent variability with period near
  $3400\,$s.} \label{fig:lc}
\end{figure*}

We extracted X-ray light curves of \piaqr from the dispersed
first-order events and found significant variability with about a
factor of two amplitude in count rate.  For the longest observation
($\sim30\ks$; Observation ID 26080) of the six segments, we also found
significant cyclic variability, with a period of about $3400\,$s.  In
Figure~\ref{fig:lc} we show light curve and hardness ratio for all
observations, and the epoch-folding period search for the segment
showing the strongest periodic signal.  To our knowledge, this is only
the second case of a clear X-ray modulation with an approximately one
hour period in a \gcas-type star, the first being found by
\citet{lopesde:2006} in HD~161103.  The other segments, which range
from $10$ to $22\ks$ exposures, had statistically significant
variability, but not as strong an indication of coherence.

The hardness ratio, defined as $(H-S)/(H+S)$, with $H$ and $S$ defined
as count rates in bands chosen to on average have equal counts 
($H:1.7$--$5.2\mang$, $S:5.2$--$13\mang$), was
relatively constant for most of the time, as we show in the left hand
side of Figure~\ref{fig:lc}, bottom panel.  The last observation,
however, stands out as extremely hard, which led us to analyze the
spectrum of that state separately.

\section{Interpretations}

The \hetg observations presented here strongly support the case that
the \piaqr B-star's companion is an accreting white dwarf.  The
emission lines of H-like ions show that there is thermally emitting
plasma, and that the temperature is high, given the lack of He-like
ions for anything but Fe.  Furthermore, the continuum is consistent
with Bremsstrahlung for the same temperature required for the emission
lines.  Even though    the spectral turnover is below $2\mang$, 
there is some leverage in the spectral curvature in the observed band.  The
temperatures we have derived are consistent with free-fall accretion
onto a white dwarf with mass between $0.5$--$0.8\msun$
\citep{mukai:2017, tsujimoto:al:2023}.  This is consistent with prior
\xmm and \nustar observations, but since emission lines are weak, the
\hetg band and spectral resolving power were required to detect and
characterize them.  The best characterized lines, those of
\eli{Mg}{12} and \eli{Si}{14}, have similar widths and centroid
offsets as in early B-star stellar winds, but they are an order of
magnitude more luminous in \piaqr, and cooler species are missing
\citep{pradhan:huenemoerder:al:2023}.  Consequently, the spectrum is incompatible
with a superposition of a powerlaw continuum with a B-star thermal spectrum.

\Change
    {
  The detection of variability with a period of about
  $3400\,\mathrm{s}$ suggests that if the companion is a white dwarf,
  that it is magnetic, such as in an
} intermediate polar where
accretion is funneled onto the white dwarf by a moderately strong
global magnetic field.  The period would then be rotational
modulation.  The breadth of the peak obtained through epoch folding is
also reminiscent of the compact object being a white dwarf \citep[see
  Figure 2 in][]{hui2012} as opposed to a neutron star where narrower
peaks are expected because of the very coherent pulsations.
\Change
    {
  The coherent variability was only seen in one observation, however.
  The others show both similar amplitude but more sporadic
  fluctuations and also changes in the mean flux level.  One would
  expect under the hypothesis of an accreting white dwarf, that if the
  accretion is steady, and the polar region of a white dwarf always
  visible, then the rotational ``clock's'' pulsations would be
  persistent.  The variations are obviously due to something more
  complicated than a coherent rotationally modulated accretion spot.
  We have evidence that very large changes can occur.  In addition to
  periodic modulation and irregular variations, there was the state
  with the high hardness ratio.  The absorption increased by an order
  of magnitude, while the flux nearly doubled, and the luminosity more
  than doubled.  We do not know if this is a random episodic or a
  phase-dependent event.  It occurred near orbital phase $0.2$, when
  viewing the companion behind the B-star, through the B-star wind,
  and perhaps downstream along the accretion column (see
  Tab.~\ref{tbl:obsinfo} and Fig.~\ref{fig:lc}).  The \swift
  monitoring by \citet{naze:rauw:smith:2019} also has their highest
  absorption near this phase.

  There is precedent for complex structure in intermediate polar
  accretion disks, with bulges, streams, and veiling which can perturb
  light curves, such as has been seen in optical, \CChange{UV, and
  X-ray observations of EX Hya \citep{belle:howell:al:2002,
  belle:howell:al:2005}.  Hiding spin modulation in X-rays with disk
  instabilities is very speculative, however, and will require longer
  term X-ray monitoring to separate coherent modulation from other
  variability.  One would have to either hide the accretion hot spot
  with strong absorption, or mask the rotational pulsations with other
  fluctuations.  Empirically, we can say that there is sometimes
  coherent X-ray modulation, but also more random fluctuations, and
  large, persistent changes in spectral hardness.  These all suggest
  complex and variable X-ray emitting and absorbing structures.
 }


  While our evidence is circumstantial, we suggest that the \piaqr
  variations indicate that a white dwarf companion is likely, given
  the episode of periodic variation.  But there is some other source
  of variability which can change emission and absorption levels.  We
  cannot determine from our dataset whether this is stochastic,
  episodic, or orbitally phase dependent.
}

\begin{figure*}[ht]
  \centering\leavevmode
  \includegraphics*[width=0.98\textwidth]{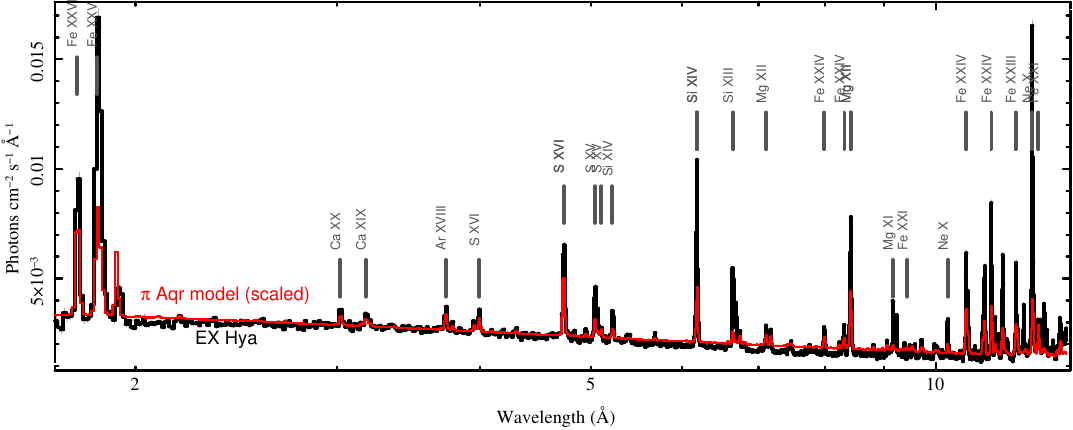} 
  \caption
      {
        The \piaqr model (red), with absorption removed and renormalized to the
        observed mean flux of EX Hya, an intermediate polar (black).  
        \CChange{The continuum shapes are similar, but the \piaqr
          emission lines are weaker relative to the continuum, and
          especially the He-like features, which have a much lower He-
          to H-like ratios.}
      } \label{fig:specexhya}
\end{figure*}

\Change {
  EX Hya is a well studied intermediate polar with 
  a white dwarf rotation period of about $4000\,$s \citep{warner:1983,
    luna:raymond:al:2015}, similar to our detected period in \piaqr.
    To compare \xray spectra, we took our  three-temperature model,
  removed the absorption and renormalized to the \hetg-observed flux
  of EX Hya.  We found overall qualitative agreement, but with
  some interesting differences.  Figure~\ref{fig:specexhya} shows our
  unabsorbed model against the observed EX Hya spectrum, which has
  little neutral absorption ($N_\mathrm{H}<10^{20}\cmmtwo$;
  \citet{luna:raymond:al:2015}).  EX Hya has emission from the H- and
  He-like ions of O, Ne, Mg, Si, S, Ar, Ca, and Fe \citep[see details
    in][]{luna:raymond:al:2015}.  In \piaqr, we only see the H-like
  lines of Mg, Si, S, and Fe, as well as He-like Fe; longward of
  $13\mang$, \piaqr absorption is too strong for detection at our
  exposure.  The differences seen, however, show that the \piaqr
  plasma is substantially hotter than that of EX Hya, since the
  He-like lines are extremely weak in \piaqr.  Our emission measure
  distribution for \piaqr, though with only 3 points, rises steadily from $\log T
  \approx 7.1$ to $8.5$.  This parallels (at about 10 times the
  amplitude) the simple cooling flow model of
  \citet{luna:raymond:al:2015} for EX Hya, which they rejected in favor of their
  empirical model which declines steeply above about
  $7.8\,$dex.  Below $\log T \sim 7.1$, if we extrapolate our emission
  measure to follow the simple cooling flow model, we cannot tell if
  such plasma exists in \piaqr because the predicted fluxes are below
  our sensitivity limit; the hot plasma greatly overwhelms any cooler
  plasma signature.  It could be that the putative white dwarf in
  \piaqr is more massive than that in EX Hya.  According to
  \citet{yuasa:nakazawa:al:2010}, a change of a few tenths of a Solar
  mass of a white dwarf can change the shock temperature by a factor
  of two.  Hence, we tentatively conclude that there is a white dwarf
  companion in \piaqr with somewhat more mass than the one in EX Hya,
  which is estimated by \citet{luna:raymond:al:2015} to be
  $0.79\msun$.  The emission measure distribution of \piaqr seems
  consistent with a cooling flow model, but higher signal-to-noise
  \xray spectra will be required to quantify the contribution of
  cooler plasma through detection of He-like ion emission.
}

Our characterization with a few-temperature, slab-absorbed model is
very rudimentary.  We are beginning to apply cooling flow models, in
which a continuous emission measure distribution is weighted inversely
by the cooling time \citep{mushotzky:al:1988, pandel:al:2005,
  bohringer:werner:2010}.
  Other refinements were implemented by
  \citet{tsujimoto:al:2023}, such as partial covering, and reflection
  models to explain Fe~K fluorescence.  The cooling flow models
  specify a mass accretion rate through the \xray luminosity and
  temperature extrema.  For a simple, slab-absorbed, geometry-independent (optically thin) cooling-flow model, 
  we obtained mass  accretion rates for the low and high hardness states of
  $4\times10^{-11}$ and $7\times10^{-11}\mdot$.  These values are
  within an order of magnitude for low-accretion-rate intermediate polars. For the expected decretion 
  disc densities at the white dwarf orbital separation \citep{jones:al:2008}, accretion rates up-to $10^{-8}-10^{-7}\mdot$ can 
  be supported (assuming spherical accretion).
  
The high-resolution \xray spectra presented here have further
supported the cases put forward by \citet{tsujimoto:al:2023} and \citet{klement:rivinius:al:2024} 
that the dominant X-ray source in \piaqr is a magnetic white dwarf, probably an
intermediate polar.  Further observations are warranted, in particular
long observations to monitor coherent variability, and phase-resolved
coverage to separate geometric from stochastic variability.

\begin{acknowledgements}

Support for this work was provided by NASA through the Smithsonian
Astrophysical Observatory (SAO) contract SV3-73016 to MIT for Support
of the Chandra X-Ray Center (CXC) and Science Instruments. CXC is
operated by SAO for and on behalf of NASA under contract NAS8-03060.
This research has made use of ISIS functions (ISISscripts) provided by
ECAP/Remeis observatory and MIT (\url{http://www.sternwarte.uni-erlangen.de/isis/}).

\end{acknowledgements}

\facility{ CXO (HETG/ACIS) }

\software{CIAO \citep{CIAO:2006},  ISIS \citep{houck2000}}


\clearpage
\bibliography{piaqr} 
\bibliographystyle{aasjournal}

\end{document}